\documentclass[12pt,preprint]{aastex}
 
\newcommand{\oversim}[2]{\protect{\mbox{\lower0.5ex\vbox{%
  \baselineskip=0pt\lineskip=0.2ex
  \ialign{$\mathsurround=0pt #1\hfil##\hfil$\crcr#2\crcr\sim\crcr}}}}}
\newcommand{\simless} {\mbox{$\,\mathrel{\mathpalette\oversim<}\,$}} 
\shorttitle{Variation of integrated star IMFs}
\shortauthors{Weidner \& Kroupa}
\begin{document}
\title{The variation of integrated star IMFs among galaxies} 
\author{C. Weidner and P. Kroupa}
\affil{Sternwarte der Universit\"at Bonn, 53121 Bonn, Germany}
\email{cweidner@astro.uni-bonn.de}
\email{pavel@astro.uni-bonn.de}

\begin{abstract}
  The integrated galaxial
  initial mass function (IGIMF) is the relevant distribution function
  containing the information on the distribution of stellar remnants,
  the number of supernovae and the chemical enrichment history of a
  galaxy.  Since most stars form in embedded star clusters with different
  masses the IGIMF becomes an integral of the assumed
  (universal or invariant) stellar IMF over the embedded
  star-cluster mass function (ECMF). For a range of
  reasonable assumptions about the IMF and the ECMF we find the
  IGIMF to be steeper (containing fewer massive stars per star) than
  the stellar IMF, but below a few $M_\odot$ it is invariant and
  identical to the stellar IMF for all
  galaxies. However, the steepening sensitively depends on the
    form of the ECMF in the low-mass regime.
  Furthermore, observations indicate a relation between the
  star formation rate of a galaxy and the most massive young stellar
  cluster in it. The assumption that this cluster mass marks the upper
  end of a young-cluster mass function leads to a connection of the
  star formation rate and the slope of the IGIMF above a few~$M_\odot$.
  The IGIMF varies with the star formation history of a galaxy.
  Notably, large variations of the IGIMF are evident for dE, dIrr
  and LSB galaxies with a small to modest stellar mass. We find
  that for any galaxy the number of 
  supernovae per star (NSNS) is suppressed relative to that expected
  for a Salpeter IMF. Dwarf galaxies have a smaller NSNS
  compared to massive galaxies.  For dwarf galaxies the NSNS varies
  substantially depending on the galaxy assembly history and the
    assumptions made about the low-mass end of the ECMF. The findings
  presented here may be of some consequence for the cosmological
  evolution of the number of supernovae per low-mass star
  and the chemical enrichment of galaxies of different mass.
\end{abstract}


\section{Introduction}
\label{sec:intro}
Over the last years it became clear that star formation takes place
mostly in embedded clusters, each cluster containing a dozen to
many million stars \citep{Kr04}. Within these clusters stars appear to
form following a universal initial mass function (IMF) with a Salpeter
power-law slope or index ($\alpha=2.35$) for stars more massive than
$1\,M_\odot$, $\xi(m) \propto m^{-\alpha}$, where $\xi dm$ is the
number of stars in the mass interval $m, m+dm$. 

\begin{figure}
\begin{center}
\vspace*{-1cm}
\includegraphics[width=8cm]{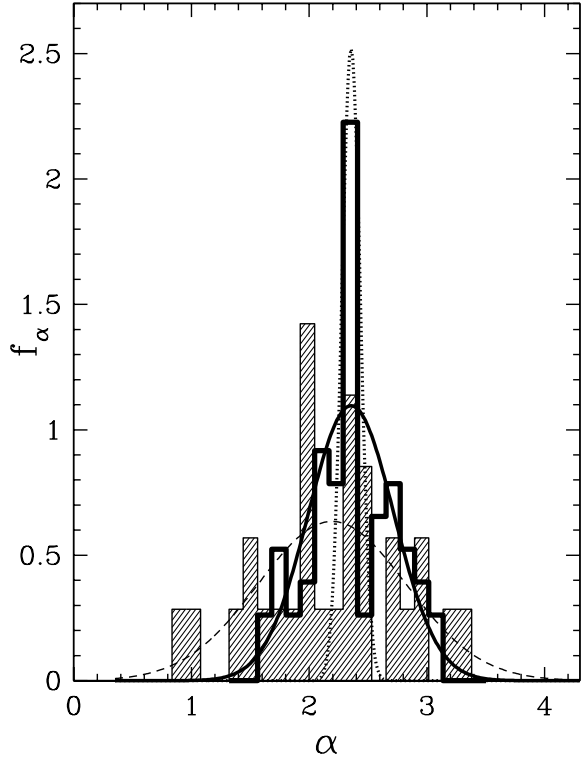}
\caption{The thick solid histogram shows the observed MF power-law
  indices $\alpha$ above 2.5 $M_{\odot}$ for an ensemble of clusters
  and associations in the MW, the LMC and SMC. Note the symmetry and
  narrowness of the empirical data that comprise a compilation of many
  OB populations in different physical systems. Note also that IMFs
  with $\alpha < 1.6$ are never observed. Such IMFs would unbind a
  young cluster rapidly due to stellar evolutionary mass-loss
  \citep{Kr01}. The shaded histogram shows 
  theoretical data: 12 star clusters with 800 to $10^{4}$ stars are set-up
  with the canonical IMF (\S~\ref{sec:maxmass}) and their slopes,
  $\alpha$, are evaluated 
  after 3 and 70 Myr of dynamical and stellar evolution. Binaries are
  merged to give the system MFs which are used to measure
  $\alpha$. The models give an even greater 
  spread than the observations although they assume a single (Salpeter) value
  for stars more massive than $0.5\,M_{\odot}$!
  Also shown are Gaussian distributions with standard deviations,
  $\sigma_{\alpha}$, obtained from the histograms. The 
  thick solid line is for the observations and the dashed for the models. The
  very narrow dotted line is the Gaussian distribution for a fixed
  $\alpha_{\rm f} = <\alpha> = 2.36$ and using only $|\alpha| \le 2
  \sigma_{\alpha}$, $\sigma_{\alpha} = 0.63$. For more details see \citet{Kr02}.}
\label{fig:alphahist}
\end{center}
\end{figure}

This has been found to be the case for a wide range of different
conditions in the Milky Way (MW), the Large and Small Magellanic
clouds (respectively LMC, SMC) and other galaxies
\citep{MaHu98,SND00,SND02,PaZa01,Mass02,Mass03,WGH02,BMK03,PBK04}.
For studies based on well-resolved stellar populations,
the observational scatter around the Salpeter value above $1
M_{\odot}$ is large but constant as a function of stellar mass
  range and consistent with a Gaussian 
distribution around this value (Fig.~\ref{fig:alphahist}). This scatter can be
explained by statistical fluctuations and stellar-dynamical
evolution of the clusters \citep{Elme99, Kr01}. The latter
changes the mass function on short and long timescales - making it
extremely difficult to measure the IMF. Careful studies taking mass
segregation into account often find Salpeter indices in young
clusters \citep[e.g. in the LMC and SMC,][]{GKK04}. Furthermore
a Salpeter IMF is found in very young clusters such as the star-burst
cluster R136 in the LMC \citep{BSB96}, NGC~1805 \citep{GGJ02},
NGC~2004, NGC~2100 \citep{GKK04} and M82-F \citep{MGV04}. The
often-observed flattening of the IMF below a few solar masses
\citep[e.g.][]{SND00} can be explained by taking mass segregation into account
\citep{SND02,DPZ04}. Especially fig.~2 from \citet{Mass03}
demonstrates the remarkable universality of the IMF over a factor of 4
in metallicity and 200 in stellar density. 

On the other hand, several observations
\citep[e.g.][]{PMS01,PMS03,SNM01,SG01,KFR03} show steeper slopes for
clusters with ages of 100 - 500 Myr and for stars ranging up to 4 -
15 $M_{\odot}$. 
For the Pleiades
cluster which has an age of about 100 Myr, \citet{MKB04} find the IMF
may be steeper than Salpeter ($\alpha\,>\,2.35$) for
$m\,\gtrsim\,2\,M_{\odot}$.
While these are important constrains, it is clear
that such clusters are already heavily dynamically evolved
\citep{KAH01}. \citet{GPE04} show for the very young Trapezium
cluster that two OB-run-away stars as far away
as about 250 pc can be traced back to it. Furthermore, about 40\% of O
stars and 5-10\% of B stars are run-away stars most probably ejected
from star-forming regions \citep{GPE04} and found up to several kpc
away from their birth places. Obviously such stars need to be
included in order to reconstruct an IMF from a present-day mass
  function (PDMF) which has not been done. But also stars
'evaporating' from a cluster after gas expulsion would travel 100 to
500 pc for clusters that are 100 to 500 Myr old even if they leave
only with a velocity of 1 $\rm km\,s^{-1}$! Thus, dynamical modelling
on a cluster-to-cluster basis would be needed to affirm possible
non-Salpeter IMFs above $1\,M_{\odot}$. 

So the case may be made that the IMF differs from cluster to cluster
\citep{Sc98,Sc04,El04a}. However, the absence of any trends with physical 
conditions together with the Gaussian distribution about the Salpeter
value above $\sim 1\,M_{\odot}$ and the similar, although somewhat
larger theoretical spread obtained for model clusters
(Fig.~\ref{fig:alphahist}), leads us to assume for now that the
stellar IMF is invariant and universal in each 
  cluster. In addition to using our default canonical Salpeter IMF
  above $1\,M_{\odot}$, we also construct models with
  $\alpha\,=\,2.70$ to account for a possibly steeper-than-Salpeter
  universal IMF above  $1\,M_{\odot}$.

Several studies show that star clusters also seem to be
distributed according to a power-law embedded
cluster mass function (ECMF), $\xi_{\rm ecl} = k_{\rm ecl} M_{\rm ecl}^{-\beta}$,
where $dN_{\rm ecl} = \xi_{\rm ecl}(M_{\rm ecl})~dM_{\rm
  ecl}$ is the number of embedded clusters in the mass interval
$M_{\rm ecl}$, $M_{\rm ecl} + dM_{\rm ecl}$ and $M_{\rm ecl}$ is
the mass in stars. For embedded stellar clusters in the solar
neighborhood with masses between 50 and 1000 $M_{\odot}$,
\citet{LaLa03} find a slope 
$\beta = 2$, while \citet{HuEl03} find $2 \simless \beta \simless 2.4
$ for $10^{3} \simless M_{\rm ecl}/M_{\odot} \simless 10^{4}$ in the
SMC and LMC, and \citet{ZhFa99} find $1.95 \pm 0.03$ for 
$10^4 \simless M_{\rm ecl}/M_\odot \simless 10^6$ in 
the Antennae galaxies. Our default assumption is
 a single-slope power-law ECMF with a lower limit of 5 $M_{\odot}$, but
 different assumptions in the low cluster-mass regime are investigated
as well.

Below roughly 100 stars ($\sim 30~M_{\odot}$) embedded
clusters dissolve within a few Myr \citep{AM01,KrBo03} but for our model only
the initial distribution of $M_{\rm ecl}$ is of interest, not the
evolution of the clusters. The detailed distribution of such small 
systems is not well known. Observational and theoretical evidence is
contradictory as \citet{SBA04} conclude that they are not major
contributors to the field, while \citet{AM01} suggest that 90\% of
star-formation takes place in such small systems. This would be
roughly equivalent to a power-law function with $\beta \sim 2.35$ down to
about 5 $M_{\odot}$ which corresponds 
to a group of about 20 stars. From a careful study of 580 open
clusters in the Milky Way \citet{FMFM04} find that 90\% of the open
clusters are formed with less than 150 stars. They are able to fit a
power-law cluster mass function as steep as $\beta = 2.7$ to the distribution
down to a cluster mass of a few solar masses. 65\% of the clusters in their
sample have reconstructed masses less than $20 M_{\odot}$. They also
conclude that 80\% of newly formed open clusters will dissolve in less
than 20 Myr. 

A universality of the ECMF
slope is not an established result as a number of works based on
HII-region luminosity functions \citep{KH89,YH99, ARR02, TWB02},
direct cluster counts \citep{CVS04} and GMC counts \citep{BR04}
indicate that it 
may vary with galaxy-type or even rather erratically. The situation
may be resolved once early cluster evolution is taken into account in
more detail; \citet{KrBo02} have already made the important point that
the cluster MF evolves rapidly within the first few $10^{6}$ yr as a
result of re-virialisation after significant residual-gas expulsion.

Under the assumption that (i) the stellar IMF is universal and
canonical and (ii) that the ECMF is also universal, we showed in
\citet{KrWe03} that the integrated galaxial initial stellar mass
function (IGIMF) must be steeper than the individual canonical IMFs in
the actual clusters.  But this steepening is critically dependent
  on the assumptions regarding the low-mass end of the ECMF.

In this contribution we develop a tool to calculate the time-depended
IGIMF for different types of galaxies. This is possible by combining
the above mentioned results with a recently discovered relation
between the maximum cluster mass formed in a star-formation epoch and
the star formation rate (SFR) of a galaxy \citep{WeKrLa04}. The
varying star formation histories (SFHs) of galaxies therefore leave
their fingerprint in the distribution of the stellar content and the
resulting chemical evolution of galaxies through a highly variable
IMF.  However, this ought to be difficult to observe because
  variations of the SFR also imply changes of the relative number
  of young and old stars, even for an invariant IGIMF.

In the next section (\S\ref{sec:meth}) we introduce the method of
calculating the IGIMF from a universal IMF, an ECMF and a star
formation history (SFH), before we present the results in
\S\ref{sec:res}, where we construct standard, maximal and minimal
models that span the range of parameters characterizing the stellar
IMF and the ECMF. The results are discussed in \S\ref{sec:conc}. Before
proceeding we note that throughout this paper IMF means the stellar
IMF which we take to be invariant.

\section{The method}
\label{sec:meth}
\subsection{The star-formation-rate--maximal-cluster-mass-relation}
\label{sec:Meclsfr}
In \citet{WeKrLa04} we derived
a relation between the maximal cluster mass in a galaxy and the current
star formation rate (SFR) of the galaxy,
\begin{equation}
\label{eq:Meclsfr}
\log_{10}(M_{\rm ecl,max}) = \log_{10}(k_{\rm ML}) + (0.75(\pm
0.03) \cdot \log_{10}{SFR}) + 6.77(\pm 0.02),
\end{equation}
where $k_{\rm ML}$ is the mass-to-light ratio, typically 0.0144 for
young ($<$ 6 Myr) clusters.  This relation connects the properties of
clustered star formation with the SFR of a galaxy. Of interest for
investigating the sensitivity of the results on the SFR is also the
use of a relation which is steeper than the default
eq.~\ref{eq:Meclsfr} but still within the three-sigma uncertainty
range of the data,
\begin{equation}
\label{eq:Meclsfr_b}
\log_{10}(M_{\rm ecl,max}) = \log_{10}(k_{\rm ML}) + (0.84 \cdot 
\log_{10}{SFR}) + 6.71 .
\end{equation}
For a given SFR this relation gives larger $M_{\rm ecl,max}$
  values than eq.~\ref{eq:Meclsfr}.

\citet{WeKrLa04} show that
eq.~\ref{eq:Meclsfr} can be reproduced theoretically for a
  power-law ECMF provided the entire population of 
star clusters with masses ranging from $M_{\rm
  ecl,min}\approx\,5\,M_\odot$ to $M_{\rm ecl,max}$ is born within a
time-span of $\delta t\approx10$~Myr, independently of the SFR.  We
refer to this time-span as a ``star-formation epoch'' of a galaxy.  As
these epochs are very short and the clusters fade quickly, a constant
$k_{\rm ML}$ can be chosen and the ECMF and the resulting IGIMF can be
determined within each epoch.
We note that we differentiate between an embedded cluster that
  contains the entire stellar population formed in one molecular cloud
  core, while on the other hand the ``initial cluster MF'' has been
  introduced as a theoretical construct arrived at by mapping
  present-day cluster masses backwards in time using classical N-body
  evolution tracks that do not take into account violent virialisation
  owing to gas expulsion \citep{KrBo02}. Thus, \citet{FMFM04}
  derive an initial cluster MF in the local MW disk with a steeper
  slope ($\beta \approx 2.7$) than $\beta \approx 2$ suggested by
  local surveys of very young embedded clusters \citep{LaLa03}. The de
  la Fuente initial cluster MF is likely to be biased towards smaller masses.

\subsection{The canonical IMF and the maximal stellar mass in a cluster}
\label{sec:maxmass}
The universal or canonical stellar IMF is conveniently written
\citep{Kr01,ReGiHa02} as a multi-power-law, $\xi(m)
\propto m^{-\alpha_{i}}$, with exponents $\alpha_0 = +0.30~(0.01 \le
m/{\rm M}_\odot < 0.08)$, $\alpha_1 = +1.30~ (0.08 \le m/{\rm M}_\odot
< 0.50)$, $\alpha_2 = +2.35~(0.50 \le m/{\rm M}_\odot < 1.00)$ and
$\alpha_3 = +2.35~(1.00 \le m/{\rm M}_\odot < 150.00)$,
ie.~\citet{Sa55} above 0.5 $M_{\odot}$. A steeper value $\alpha_3=2.7$ has
been suggested by \citet{KTG93} based on Scalo's
(1986) star-count analysis of the relatively local Galactic field.
Such a steep IMF above $1\,M_\odot$ may be the correct IMF resulting
from star-cluster formation if corrections for unresolved multiples
among massive stars are applied \citep{SaRi91,MaZi01}.
  Alternatively, the Scalo value may reflect a steeper IGIMF
  \citep{KrWe03}. We use both, the canonical Salpeter $\alpha_3$ and
the Scalo value. 

The cluster mass in stars is
\begin{equation}
\label{eq:Mint}
M_{\rm ecl} = \int_{m_{\rm low}}^{m_{\rm max}}m \cdot \xi(m)~dm,
\end{equation}
where $m_{\rm low}=0.01\,M_\odot$ is the adopted minimum mass given by
opacity-limited fragmentation while $m_{\rm max}$ is the maximum
stellar mass that can occur in the cluster.  Apart from not well
understood physical reasons \citep{WeKr04}, it cannot be arbitrarily
high because the 
cluster mass sets limits on its value. These can be formulated through
conditional statistics. Statistically, a cluster with mass $M_{\rm
  ecl}$ contains exactly one most massive star,
\begin{equation}
1 = \int_{m_{\rm max}}^{m_{\rm max*}}\xi(m)~dm,
\end{equation}
The resulting equation, $m_{\rm max}(M_{\rm ecl})$, cannot be solved
analytically. \citet{WeKr04} solve it numerically and show that an
upper bound on $m_{\rm max}\le m_{\rm max*}\approx 150\,M_\odot$ must
exist as otherwise there would be too many stars with masses larger
than about $100\,M_\odot$ in the LMC star-burst cluster R136.
\citet{WeKr04} interpret $m_{\rm max*}$ to be a fundamental upper
stellar mass. Intriguingly, the same stellar mass limit is noted
  by \citet{Fi05} for the metal-rich MW-nuclear cluster Arches
  suggesting that perhaps $m_{\rm max*} \approx 150\,M_{\odot}$ may be
quite independent of metallicity, and \citet{OC05} find
  such an upper mass limit based on statistical examinations of
  several clusters. Numerical simulations of star-formation in clusters 
\citep{BVB04} also indicate that the mass of the most massive star
scales with the system (cluster) mass and is not a purely random value
but a conditional one.

\subsection{The integrated galaxial initial mass function}
\label{sec:IGIMF}
Under the assumption of an invariant canonical stellar IMF in
  clusters and an invariant ECMF (\S~\ref{sec:intro}),
the integrated galaxial initial mass function (IGIMF) is calculated by
adding all stars in all clusters \citep[as already noted by][]{VB82, VB83}, 
\begin{equation} 
\label{eq:igimf}
\xi_{\rm IGIMF}(m,t) = \int_{M_{\rm ecl,min}}^{M_{\rm ecl,max}(SFR(t))}
\xi(m\le m_{\rm max})~\xi_{\rm ecl}(M_{\rm ecl})~dM_{\rm ecl}. 
\end{equation}
Thus $\xi(m\le m_{\rm max})~\xi_{\rm ecl}(M_{\rm ecl})~dM_{\rm
    ecl}$ is the stellar IMF contributed by $\xi_{\rm ecl}~dM_{\rm ecl}$
  clusters with mass near $M_{\rm ecl}$.
While $M_{\rm ecl,max}$ follows from eq.~\ref{eq:Meclsfr}, $M_{\rm
  ecl,min}\,=\,5\,M_{\odot}$ is adopted.  The stellar mass in an
embedded cluster is $M_{\rm ecl}$ so that the mass in stars and gas of
the whole embedded cluster amounts to $M_{\rm ecl} / \epsilon$ for a star
formation efficiency of $\epsilon$ \citep[$\approx
1/3$,][]{LaLa03}. Note that in 
\citet{KrWe03} we referred to eq.~\ref{eq:igimf} as the ``field-star
IMF'', $\xi_{\rm field}$. This is strictly speaking not correct,
because the IGIMF includes all stars in all clusters {\it and} the galactic
field which consists of already dispersed clusters.  However, as long as the
surviving and newly-born star-clusters do not constitute a significant
stellar contribution to the whole galaxy and ignoring issues
  arising from a flux limit, $\xi_{\rm IGIMF}\approx \xi_{\rm
  field}$ for the time-averaged galaxy. 
Another way of looking at eq.~\ref{eq:igimf} is to consider joint
probabilities: The joint probability for finding a star of mass
$m$ in a cluster of mass $M_{\rm ecl}$ that has an upper
stellar mass limit $m_{\rm max}$ is $P(m,M_{\rm ecl}) = P(m|M_{\rm
  ecl}) \times P(M_{\rm ecl})$, where $M_{\rm ecl} = M_{\rm
  ecl}(m_{\rm max})$. Integrating over all cluster masses and scaling
to the correct number of stars then recovers eq.~\ref{eq:igimf}.

The complete procedure is implemented in the following way: After the
specification of a SFR for an individual star formation epoch the
resulting $M_{\rm ecl, max}$ (eq.~\ref{eq:Meclsfr}) is used to
construct an ECMF with a predefined slope $\beta$. Each individual
cluster in the ECMF is constructed from a predefined IMF up to a
stellar mass limit determined by the mass of the individual cluster
($m_{\rm max}(M_{\rm ecl})$). Then the separate IMFs of the clusters
are added-up to give the IGIMF for this epoch (eq.~\ref{eq:igimf}).
This is repeated with different SFRs to account for a varying SFH
until the desired mass of the galaxy is reached at the desired age.
The IGIMFs of the individual epochs are added to give the final IGIMF
of the model galaxy.

It should be noted here that the IGIMF is not the PDMF as stellar
evolution is not included, but it is the galaxy-wide IMF (galaxial
IMF) which may be used to estimate certain properties of galaxies (as in
\S~\ref{sec:SNrates}). The final IGIMF is, strictly speaking, only a
theoretical construct because it counts all massive stars
irrespective of when they are formed. To quantify the stellar population
at any given moment we would need to include stellar evolution.

\section{Results}
\label{sec:res}
Here we discuss the implications of our model for three different
  cases. In the first scenario (\S~\ref{sec:1}) so-called 'standard'
  parameters are used (stellar IMF slope above
  $0.5~M_{\odot}$ being $\alpha_{2}\,=\,\alpha_{3}\,=\,2.35$, and an ECMF slope 
  $\beta\,=\,2.35$ for 5 $M_{\odot} \le M_{\rm ecl}$, taking $\beta =
  \alpha_{3}$ for simplicity). In \S~\ref{sec:2} we
  investigate a parameter set within the allowed range but which
  maximizes the effect on the IGIMF
  ($\alpha_{3}\,=\,2.70,\,\beta\,=\,2.35$ for 5 $M_{\odot} \le M_{\rm 
    ecl}$). Finally a set of parameters out of the allowed ones that
  minimize the effect is studied in
  \S~\ref{sec:3} ($\alpha_{3}\,=\,2.35$ and a two-part power-law ECMF
  with $\beta_{1}\,=\,1$ for $5 \le 
  M_{\rm ecl}/M_{\odot}\,\le\,50,\,\beta_{2}\,=\,2$ for $50 M_{\odot}\,\le\,M_{\rm
    ecl}$), with some results also given for a ECMF truncated at 50
  $M_{\odot}$ with $\beta = 2$ for $M_{\rm ecl}\,\ge\,50\,M_{\odot}$.

\subsection{The 'standard' scenario}
\label{sec:1}
\subsubsection{The IGIMF}
\label{sec:rIGIMF}
\begin{figure}
\begin{center}
\vspace*{-1cm}
\includegraphics[width=8cm]{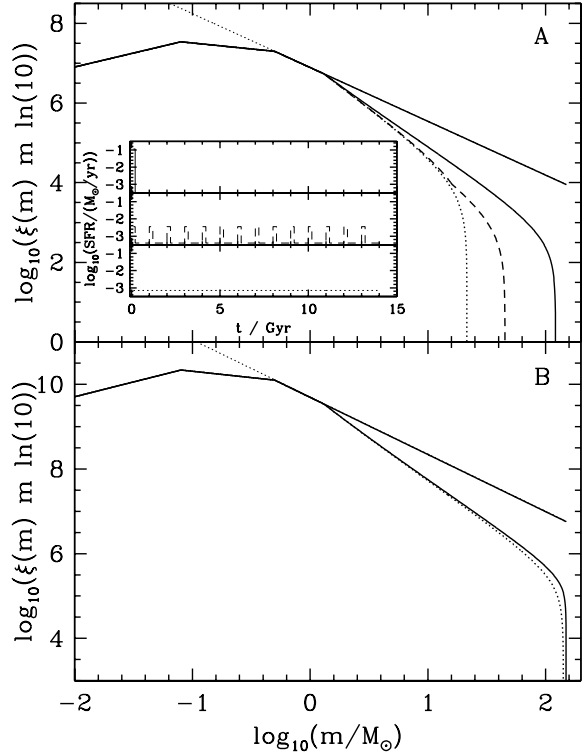}
\caption{{\it Panel A:} IGIMFs for three galaxies with final stellar
  masses of $10^{7}\,M_{\odot}$ but different SFHs. The {\it solid
    curve} results from a single 100 Myr long 
  burst of star formation. The {\it dashed curve} assumes a periodic
  SFH with 14 peaks each 100 Myr long and 900 Myr quiescent periods in
  between.  The {\it thick dotted curve} assumes a constant SFR over
  14 Gyr. For all cases the canonical IMF power-law 
  slope above $1\,M_{\odot}$ ($\alpha_{3}=2.35$) is used and the ECMF
  slope is $\beta=2.35$. The {\it straight solid line} above 0.5
  $M_{\odot}$ shows the canonical input IMF for comparison. The {\it
    thin dotted line} is a Salpeter IMF extended to low mass. Note the
  downturn of the IGIMFs at high masses. It results from the inclusion
  of a limiting maximum mass, $m_{\rm max}\le m_{\rm max*}\,=\,150
  M_{\odot}$, into our 
  formalism \citep{WeKr04,Fi05}. {\it Panel B:} As {\it panel A} but for a
  galaxy with $10^{10}\,M_{\odot}$ in stars. The IGIMF is 
  only shown for an initial-burst SFH ({\it solid curve}), giving an
  E-type galaxy, and for a continuous SFH ({\it dotted curve}) giving
  a MW-type galaxy.}
\label{fig:1e0710a235}
\end{center}
\end{figure}

{\it Panel A} of Fig.~\ref{fig:1e0710a235} shows the large differences
of the final IGIMF for a galaxy with $M_{\rm gal} \simless
10^{9}\,M_{\odot}$ in stars. Depending on the SFH such 
a galaxy can be a dwarf spheroidal (dSph), a dwarf elliptical (dE), a
dwarf irregular (dIrr) or a low-surface-brightness galaxy (LSB). We
refer to such galaxies as ``dwarf'' galaxies. While in a single SF burst
resulting in a dE system the IGIMF is populated up to the highest
stellar masses, in the case of a LSB galaxy with a constant SFR the
slope is much steeper and only stars up to $\sim 25\,M_{\odot}$ ever
form. Low-mass LSB galaxies thus appear chemically very young.

Considering more massive galaxies ($M_{\rm gal}>\,10^{10}\,M_{\odot}$
in stars, {\it panel B} of Fig.~\ref{fig:1e0710a235}), the variation
of the IGIMF with the 
SFR is not as pronounced as before. The weak dependence on the SFH
comes about because even the continuous SFR is high enough to sample the
ECMF to massive clusters such that the massive stars end up being
well-represented. Note however that the resulting IGIMF is always
significantly steeper than the canonical IMF, and equal to the
canonical IMF below a few~$M_\odot$ \citep{KrWe03}.

\begin{figure}
\begin{center}
\vspace*{-1cm}
\includegraphics[width=8cm]{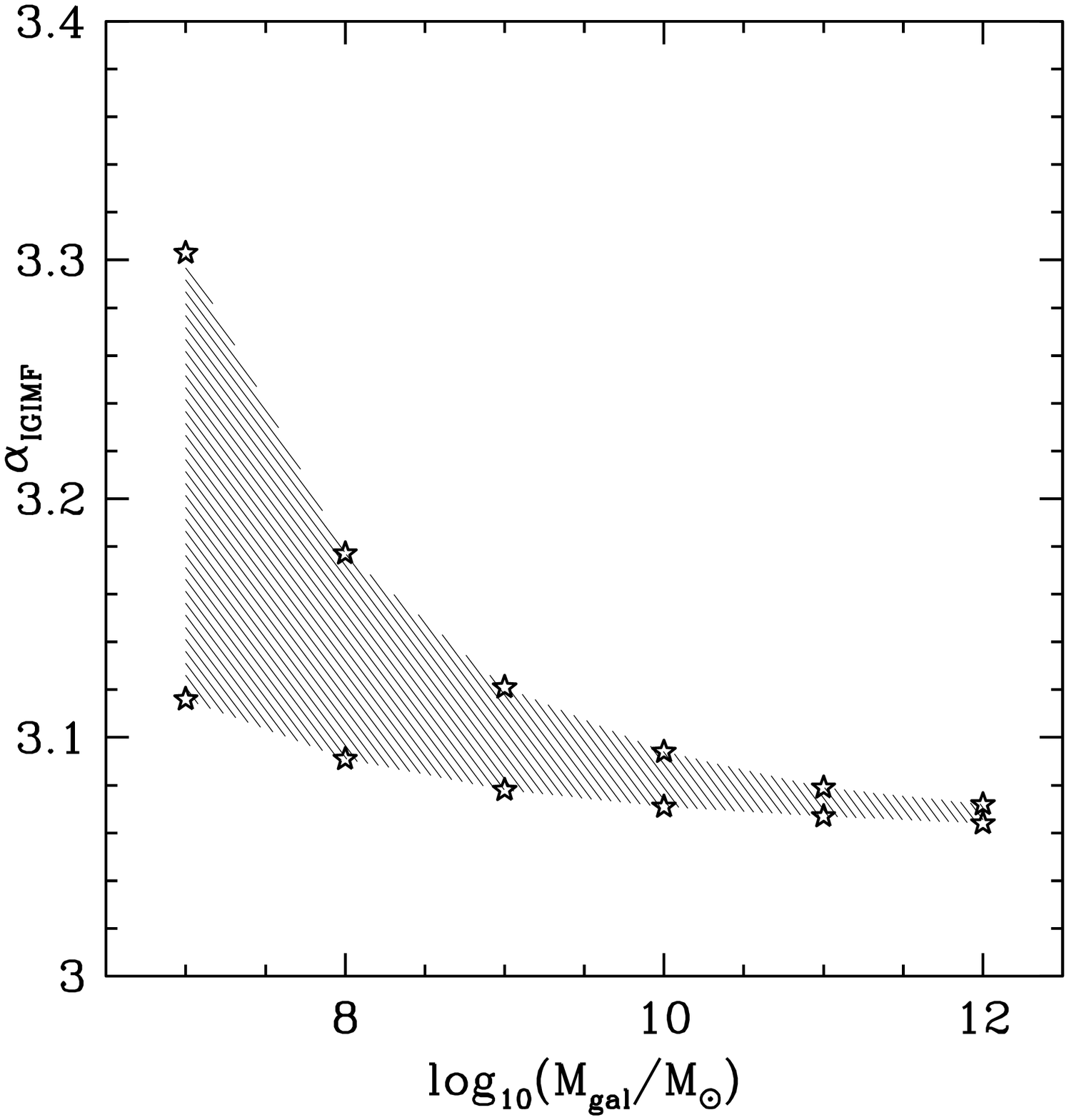}
\caption{The resulting IGIMF slopes above $m_{\rm knick}\,=\,1.3\,M_{\odot}$ are
  shown in dependence of stellar galaxy mass for different models, as
  indicated.  The lower bound of the shaded area is for an initial
  SF burst forming the entire stellar galaxy while the
  upper bound is derived for a continuous SFH.
  The symbols correspond to calculated models. The region
  within the shaded area corresponds to models with prolonged and
  time-varying SFHs. Note that we classify galaxies with $M_{\rm gal} \le
  10^{9}\,M_{\odot}$ as dwarfs. These include dSph, dE, dIrr and LSB
  galaxies. A Salpeter IMF has $\alpha\,=\,2.35$.}
\label{fig:massslc1a}
\end{center}
\end{figure}

\begin{figure}
\begin{center}
\vspace*{-1cm}
\includegraphics[width=8cm]{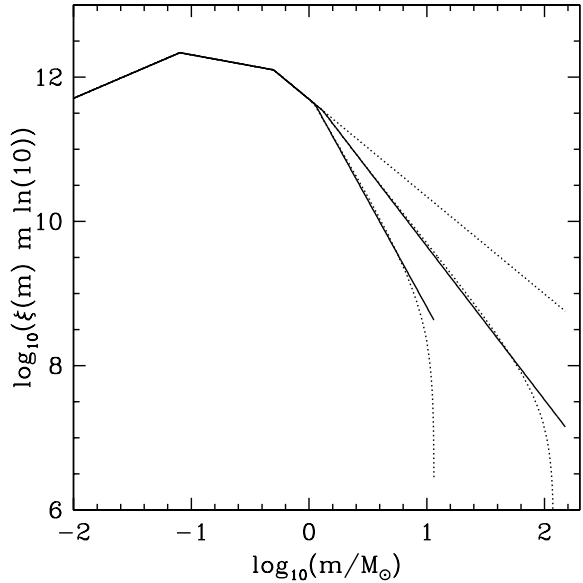}
\vspace*{-2cm}
\caption{Two examples of how $\alpha_{\rm IGMF}$ is fitted. The straight
dotted line is the canonical IMF (with $\alpha_{3}\,=\,2.35$)
while the other dotted lines are resulting IGIMFs for two models with
$M_{\rm gal}\,=\,10^{7}\,M_{\odot}$. The left one has an input
$\alpha_{3}\,=\,2.70$ and continuous star-formation over 14 Gyr and the
right one an input $\alpha_{3}\,=\,2.35$ and is formed in a single 100 Myr
burst of star-formation. In both cases the ECMF has a slope of
$\beta\,=\,2.35$. The solid lines show the fits derived from our
algorithm based on the assumption that the number of stars beyond
$m_{\rm knick}$ is equal for the corresponding solid and dotted lines.}
\label{fig:figfit}
\end{center}
\end{figure}

The variation of the IGIMF is best illustrated by comparing the slope, 
$\alpha_{\rm IGIMF}$, of the resulting IGIMFs
(Fig.~\ref{fig:massslc1a}). In order to fit IGIMF slopes consistently 
the number of stars above $m_{\rm knick}$ (see
Tab.~\ref{mknick}) is calculated for each IGIMF up to the maximal mass
of the corresponding model. This number, $N_{1}$, is compared with the
corresponding number $N_{2}$ calculated for a representative IGIMF
with a single slope $\alpha_{\rm IGIMF}$ above $m_{\rm knick}$ up to
the same mass limit. That $\alpha_{\rm IGIMF}$
is chosen to represent the IGIMF for which $N_{1}\,=\,N_{2}$.
Fig.~\ref{fig:figfit} shows an example of the results of the automated fitting
routine. 

For dwarf galaxies a large difference is seen 
between models with continuous star formation (upper bound of the
shaded area in Fig.~\ref{fig:massslc1a}) and a single, initial-burst
of star formation (lower bound of the shaded area). All other SFHs
would produce results in-between these two extremes.

\begin{table}[h]
\centering
\caption{\label{mknick} Down-turn (``knick'') masses for different
  model assumptions. For e.g., $m_{\rm knick}$ = 1.307 for the
  canonical models shown in Fig.~\ref{fig:1e0710a235}. $\beta_{1}$ and
$\beta_{2}$ are the power-law indices of the ECMF below and above
$50\,M_{\odot}$, respectively.}
\vspace*{0.5cm}
\begin{tabular}{ccccc}
\hline
$\alpha_{3}$&$\beta_{1}$&$\beta_{2}$&$M_{\rm ecl,min}$&$m_{\rm knick}$\\
&&&[$M_{\odot}$]&[$M_{\odot}$]\\
\hline
2.35&2.35&2.35&5&1.307\\
2.70&2.35&2.35&5&1.120\\
2.35&1.00&2.00&5&1.307\\
2.35&2.00&2.00&5&1.307\\
2.35&2.35&2.35&50&5.560\\
2.35&2.35&2.35&100&8.823\\
\hline
\end{tabular}
\end{table}

\subsubsection{Number of supernovae of type II}
\label{sec:SNrates}
\begin{figure}
\begin{center}
\includegraphics[width=8cm]{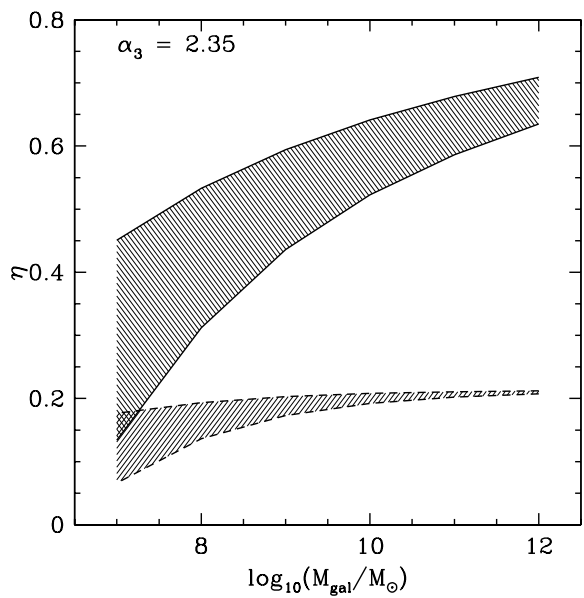}
\caption{The relative NSNS in dependence of the stellar mass of the
  galaxy. The {\it upper shaded area} represents models with an ECMF
  slope $\beta\,=\,2$, and for the {\it lower area} $\beta\,=\,2.35$.
  The upper limit for each shaded region corresponds to a single-burst
  model while the lower limit is for continuous SF models.
  The (input) IMF slope for stars above $1\,M_{\odot}$ is $\alpha_{3}\,=\,2.35$
  (canonical) and the $\eta$ plotted here are relative to the
  $\rm NSNS_{\rm can}$ calculated for a constant canonical IMF (Salpeter above
  $0.5\,M_{\odot}$).} 
\label{fig:rates235}
\end{center}
\end{figure}

The steeper IGIMF slopes imply a less-frequent occurrence of type II
supernovae (SN) in galaxies.  To quantify this we calculate from the
IGIMFs of each galaxy model the total number of stars formed above
$8\,M_{\odot}$ over a period of 14 Gyr. We then divide this number by
the total number of stars in the galaxy in order to get the number of
SN per star, $\rm NSNS_{\rm IG}$ (integrated galaxial NSNS). The $\rm
NSNS_{\rm IG}$ is then divided by the NSNS for models in 
which the same mass in stars is distributed according to the canonical
IMF ($\alpha_{3}\,=\,2.35$ for $m\,>\,1.0\,M_{\odot}$) to get 
the {\it relative NSNS for the IGIMF models in comparison to the NSNS
  expected from applying the (incorrect) canonical IMF} containing
stars between 0.01 and 150 $M_{\odot}$. For this (incorrect) model in
which the stellar upper mass limit does not depend on the SFR, the
$\rm NSNS_{\rm can}$ = 0.003374 SNII per star. Thus, we have 
\begin{equation}
{\rm
NSNS_{\rm IG}} = \frac{\int_{8}^{m_{\rm max*}} \xi_{\rm
    IGIMF}(m)\,dm}{\int_{0.01}^{m_{\rm max*}} \xi_{\rm IGIMF}(m)\,dm},
\end{equation}
and likewise for $\rm NSNS_{\rm can}$ where $\xi_{\rm IGIMF}$ is
replaced by $\xi(m)$. The relative NSNS is then given by,
\begin{equation}
\eta = \frac{\rm NSNS_{\rm IG}}{\rm NSNS_{\rm can}}.
\end{equation}
The resulting $\eta$ values are shown in Fig.~\ref{fig:rates235}
for $\beta$ = 2 and 2.35. 

Two main effects are visible. Firstly the $\eta$ are always
smaller than for models with a constant canonical IMF ($\eta$
$<1$).  For example, for galaxies with a stellar mass of
$10^7\,M_\odot$ the actual NSNS would be only 10~per cent of the NSNS
expected traditionally by adopting a canonical IMF. Secondly, there is
a strong dependence on galaxy mass: $\eta$ decreases substantially
with decreasing mass in stars.

The NSNS may thus be a strong function of the cosmological epoch,
given that present-day galaxies are build-up 
from dwarfs through hierarchical merging.

Note that the traditional calculation based on an
invariant Salpeter or canonical IMF leads to a constant NSNS
independent of the SFR or galaxy mass. Often a Salpeter IMF is used for
stars between $0.1$ and $100\, M_{\odot}$. Such
an IMF has the constant $\rm NSNS_{\rm Salp}$ = 0.002512 which is 1.3
times smaller than the above constant $\rm NSNS_{\rm can}$ = 0.003374. 

\subsection{The 'maximal' scenario}
\label{sec:2}
\begin{figure}
\begin{center}
\vspace*{-1cm}
\includegraphics[width=8cm]{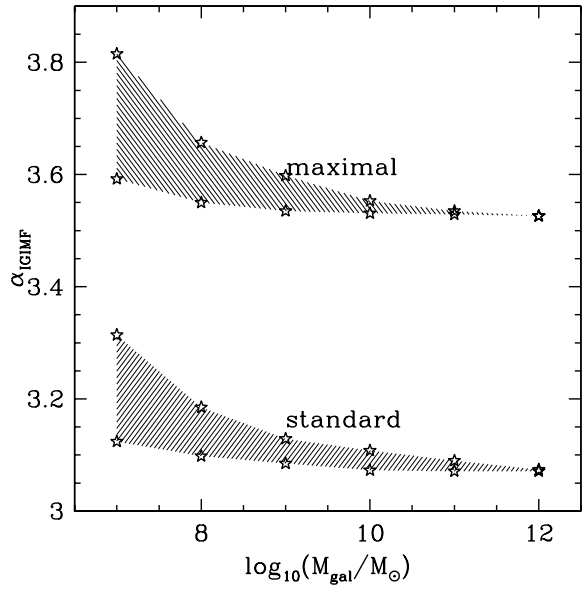}
\caption{As Fig.~\ref{fig:massslc1a} but with $\alpha_{3}=2.70$. The
  results of the 'standard' scenario with $\alpha_{3}\,=\,2.35$ are
  the lower hatched region.} 
\label{fig:massslc1b}
\end{center}
\end{figure}

In this subsection we explore our so-called 'maximum' scenario where we
adopt a slope of $\alpha_{3} = 2.7$ for the stellar IMF above
$1.0\,M_{\odot}$. Again we find that
the effect is larger for less-massive galaxies, and that the
dependence of the IGIMF on the SFR is weak for massive galaxies. 
Fig.~\ref{fig:massslc1b} illustrates the slopes of the IGIMF
above $m_{\rm knick}$ in the maximal and the standard
scenario introduced in \S~\ref{sec:1}. Due to the steeper input slope
all resulting slopes are shifted to larger values by about 0.5 dex and
the spread between single initial-burst 
models and continuous SF models is somewhat larger for dwarf galaxies
than in the standard scenario.

\subsubsection{Number of supernovae of type II}
\label{sec:SNrates2}
\begin{figure}
\begin{center}
\includegraphics[width=8cm]{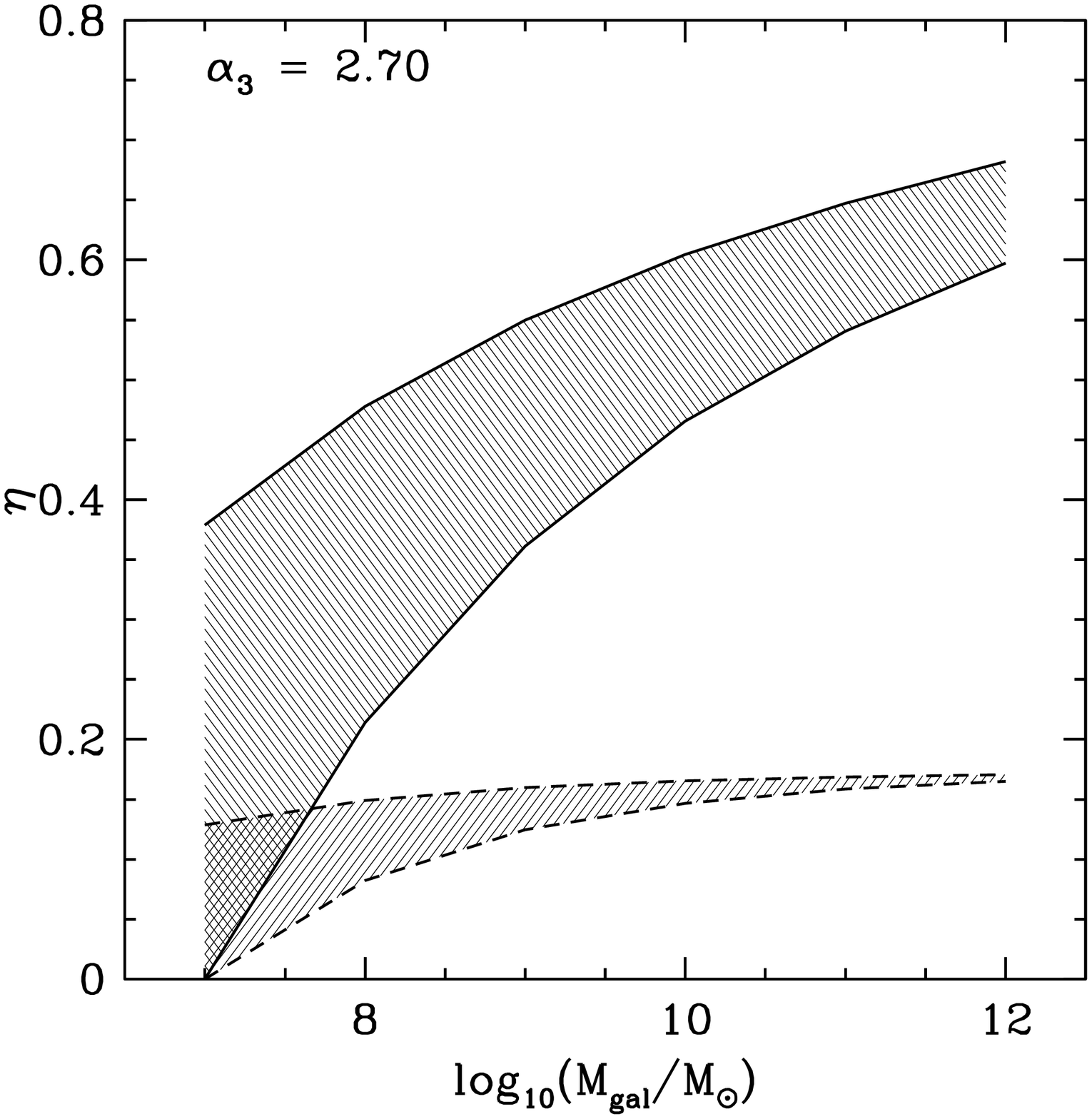}
\caption{Like Fig.~\ref{fig:rates235} but with an (input) IMF
  slope of $\alpha_{3} = 2.70$ for stars above $1\,M_{\odot}$.}
\label{fig:rates270}
\end{center}
\end{figure}

The relative NSNS for the maximal scenario are shown in
Fig.~\ref{fig:rates270}, again for two cases of the ECMF slope
$\beta$. As expected $\eta$ drops in comparison to the standard
scenario (Fig.~\ref{fig:rates235}), and in 
the case of low-mass galaxies $\eta$ can even become zero, meaning
that there would be no SNII in such galaxies if the SFR is low enough. 

\subsection{Other ECMF and the 'minimal' scenario}
\label{sec:3}
\begin{figure}
\begin{center}
\vspace*{-1cm}
\includegraphics[width=8cm]{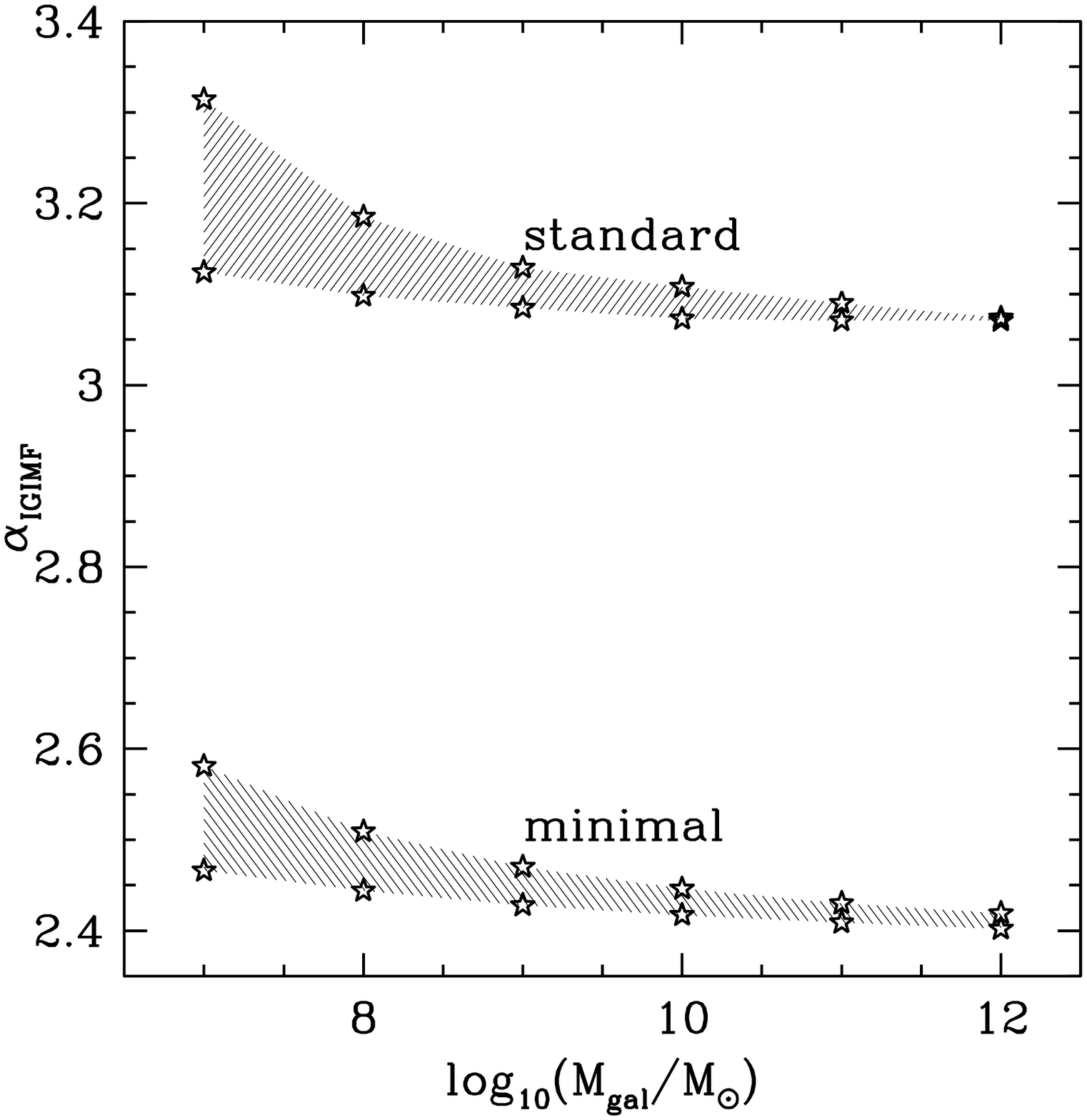}
\caption{As Fig.~\ref{fig:massslc1a} but with a 2-part-power-law ECMF
  with $\beta_{1}\,=\,1.00$ and $\beta_{2}\,=\,2.00$ and $\alpha_{3}=2.35$. The
  results of the 'standard' scenario are the upper hatched region with
  $\beta\,=\,2.35$ and $\alpha_{3}\,=\,2.35$.}
\label{fig:maslopemin}
\end{center}
\end{figure}

In order to investigate the sensitivity of our results on a non-default
ECMF at the lower mass end, three different assumptions for the ECMF
are used:~First by a flattening below 50 $M_{\odot}$ to
a slope of $\beta_{1}$ = 1 with $\beta_{2}$ = 2.35 above 50
$M_{\odot}$, and secondly by a lower-mass cut-off in the ECMF at 50
$M_{\odot}$ with $\beta$ = 2.35 above. As a result, the down-turn,
$m_{\rm knick}$,  
of the galaxial IMF is shifted to higher masses, but the deviations
from the Salpeter value above 1 $M_{\odot}$ remain large.
Finally we use a lower mass slope for the ECMF of
$\beta_{1}$ = 1 below $50~M_{\odot}$ and additionally change the slope above to
$\beta_{2}$ = 2 in order to explore a minimal case. Due to the smaller
$\beta_{2}$ value of 2 in comparison with our other models the
deviation from the canonical value is small in both panels of
Fig.~\ref{fig:maslopemin} for the initial-burst models and for
massive galaxies in general. Never the less, for dwarf galaxies the
difference between continuous SF and an initial-burst remains substantial.

\subsubsection{Number of supernovae of type II}
\label{sec:SNrates3}

\begin{figure}
\begin{center}
\includegraphics[width=8cm]{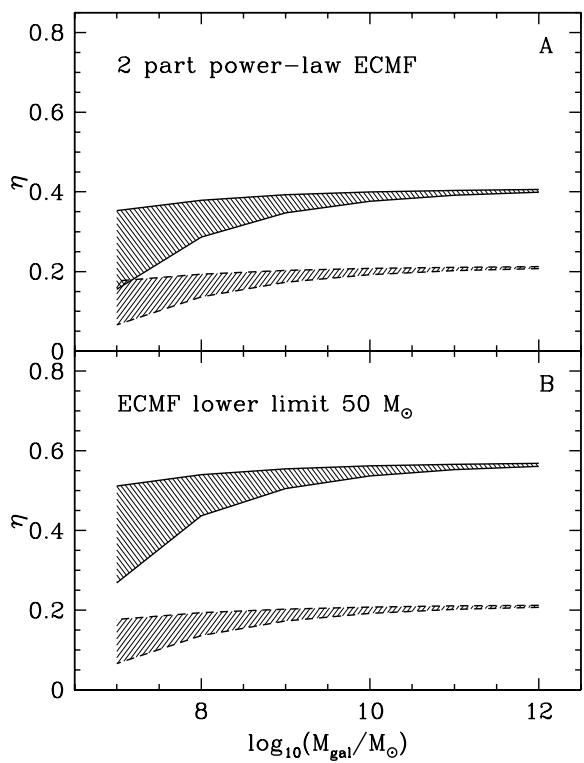}
\caption{
  {\it Panel A}: Like Fig.~\ref{fig:rates235} but with a
  flattening to $\beta_{1} = 1$ of the ECMF below 50 $M_{\odot}$. {\it
    Panel B}: Like {\it panel 
    A} but with $M_{\rm ecl,min} =~50~M_{\odot}$ for the ECMF. In both
  cases $\beta_{2}\,=\,2.35$ for $M_{\rm ecl}\,>\,50\,M_{\odot}$ and the
  shaded area enclosed by the {\it dashed curves} are  
  the standard values from Fig.~\ref{fig:rates235} (for
  $\alpha_{3}\,=\,2.35$ and $\beta\,=\,2.35$).} 
\label{fig:ratesPS}
\end{center}
\end{figure}

\begin{figure}
\begin{center}
\includegraphics[width=8cm]{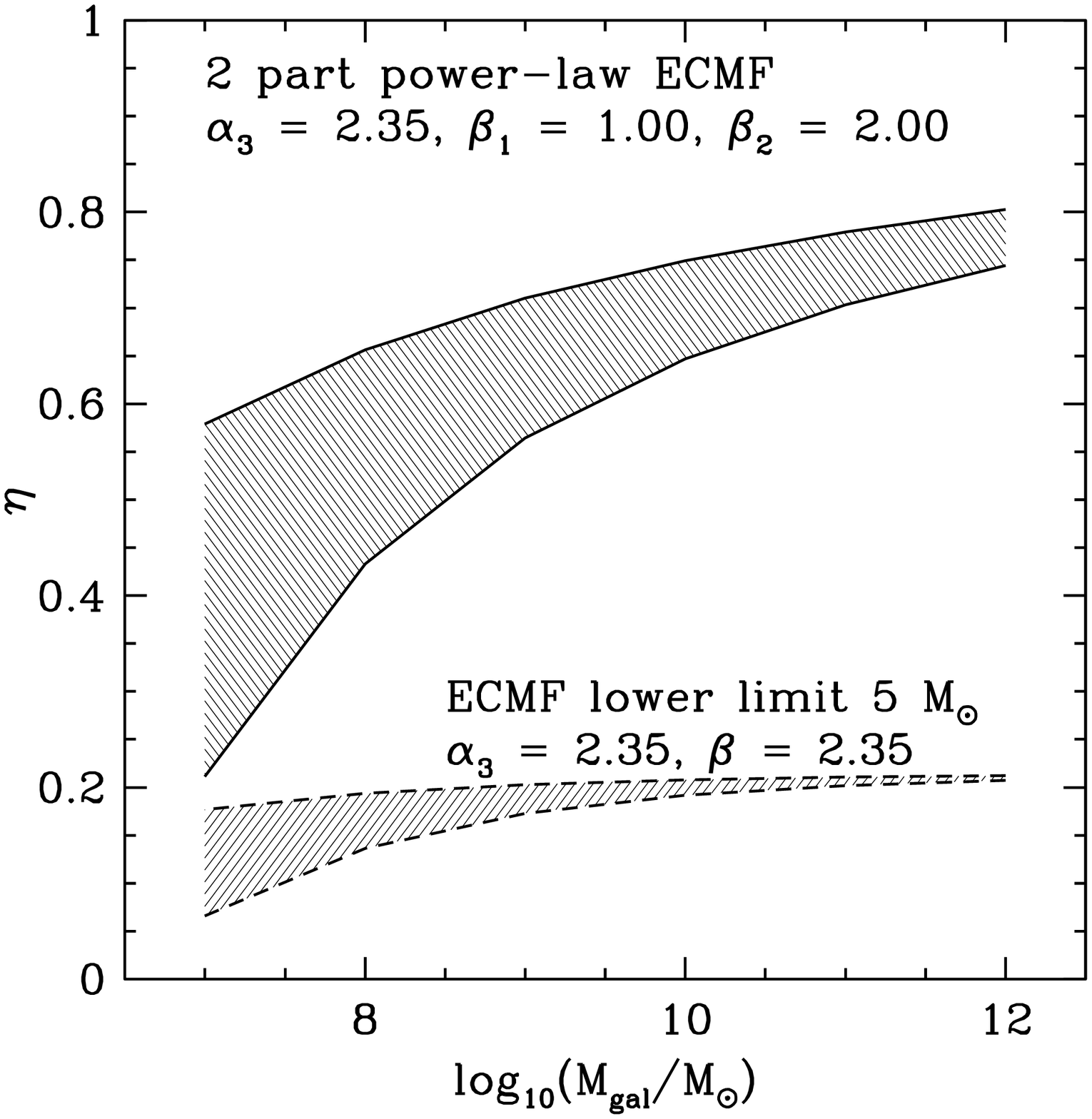}
\caption{The minimal scenario, like Fig.~\ref{fig:rates235} but with
  an ECMF slope 
  $\beta_{1}$ = 1 below $50~M_{\odot}$ and $\beta_{2} = 2$ for $M_{\rm
    ecl}\,>\,50\,M_{\odot}$. The shaded area enclosed by the {\it
    dashed curves} are the standard values from
  Fig.~\ref{fig:rates235} (for $\alpha_{3}$ = 2.35 and $\beta$ = 2.35).} 
\label{fig:ratesb200l}
\end{center}
\end{figure}

In Fig.~\ref{fig:ratesPS} $\eta$ is shown for a flat
($\beta_{1}\,=\,1$) ECMF below 50 $M_{\odot}$ ({\it panel A}) and for a
cut-off below 50 $M_{\odot}$ ({\it panel B}). While in the first case
$\eta$ is still very low (20 to 40\% of the corresponding canonical
value), in the second case $\eta$ increases up to 60\% thus reducing the
effect. In Fig.~\ref{fig:ratesb200l} is $\eta$ shown for the
  minimal model with an ECMF slope of $\beta_{1}\,=\,1$ below 
  $50\,M_{\odot}$ and $\beta_{2}\,=\,2$ above $50\,M_{\odot}$. Here the
  deviations from the invariant canonical model are only about 20\% for massive
  galaxies but are still about 50\% for dwarf galaxies.
Thus we find that even for assumptions that minimize the effects due
to clustered star formation, $\eta < 0.8$ for all
galaxies, with smaller values for less-massive galaxies.

\subsection{Comparison of the scenarios}
\label{comparison}
\begin{figure}
\begin{center}
\includegraphics[width=8cm]{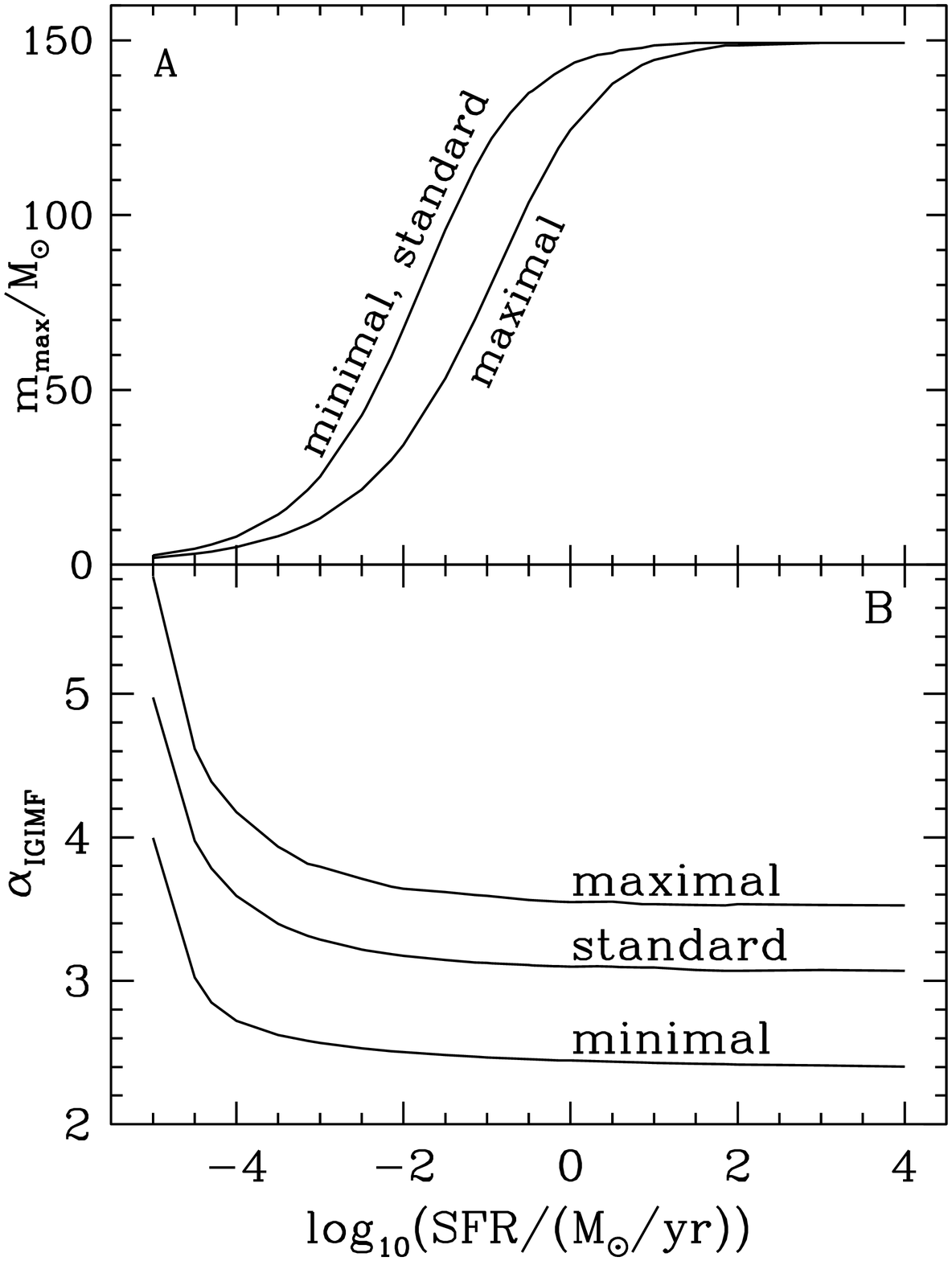}
\caption{
  {\it Panel A}: The upper stellar mass limit ($m_{\rm
    max}\,\le\,m_{\rm max,*}\,=\,150\,M_{\odot}$) as a
  function of the SFR of a galaxy per epoch of 10 Myr duration for the
  three scenarios. While for SFRs 
  above $1 M_{\odot}\,{\rm yr}^{-1}$ the IMF is sampled 
  up to the maximum stellar mass limit of about 150 $M_{\odot}$
  \citep[for details see][]{WeKr04}, the limit declines for lower
  SFRs. {\it Panel B}: The slope of the IGIMF generated during a
  star-formation epoch as a function of the SFR during that epoch.} 
\label{fig:maxmasfr}
\end{center}
\end{figure}

For the efficient construction of models (IGIMF vs SFH) and for a
useful comparison with observations we also compute the upper stellar
mass limit and the IGIMF slope as functions of the galaxial SFR per
epoch of 10 Myr duration (\S~\ref{sec:Meclsfr}) for all three
scenarios. The results are presented in Fig.~\ref{fig:maxmasfr}.  The
upper mass limits ({\it panel A} of Fig.~\ref{fig:maxmasfr}) are equal
for the minimal and the standard scenario. The slopes of the IGIMF
({\it panel B} of Fig.~\ref{fig:maxmasfr}) span a large range between
the different scenarios, but in all cases steeper slopes are to be
expected for low SFRs. The data plotted in Fig.~\ref{fig:maxmasfr}
allow the construction of IGIMFs without the need to perform the
detailed modelling described in this paper.

It is important to note
here that the SFRs in Fig.~\ref{fig:maxmasfr} are not the averaged
SFRs of the galaxy but the SFRs during the star-formation epoch. Thus,
for example, a galaxy with $M_{\rm gal} = 10^{7}\,M_{\odot}$
produced continuously over 1~Gyr has the same SFR of
$10^{-2}\,M_{\odot}\,{\rm yr}^{-1}$ and the same IGIMF and $\eta$ as a
galaxy with $M_{\rm gal} = 14 \times 10^{7}\,M_{\odot}$ produced
continuously over 14~Gyr. In 
order to get the complete stellar population of a galaxy the IGIMF
slope of each star-formation epoch of the SFH has to be calculated,
resulting in a series of populations, each with its own $\xi_{\rm
  IGIMF}(m,t)$.  These have to be added up to form the complete galaxy
with the overall $\xi_{\rm IGIMF}$.

\section{Conclusions and Discussion}
\label{sec:conc}
We show that it is possible to explain varying stellar populations in
galaxies by a simple mechanism based on universal principles for all
galaxies. Our assumptions are (i) that all stars are born in clusters
following a universal embedded cluster mass function which is
populated up to a maximum cluster mass which depends on the star
formation rate of a galaxy, and (ii) that within these clusters all
stars are born from a universal stellar IMF. But the magnitude of
  the effect strongly depends on the shape of the ECMF for low-mass
  clusters. The combination of a varying SFR with the empirical
$M_{\rm ecl,max}(SFR)$ relation, together with the stellar mass being
limited by the cluster mass, naturally leads to a time-dependent
integrated galaxial initial stellar mass function. This IGIMF also
depends on the mass of the galaxy by virtue of the average level of
the SFR being proportional to $M_{\rm gal}$. The steep IGIMF and
  the variations -- especially for dwarf galaxies -- may have
  implications for the chemo-dynamical evolution of galaxies. But we
  note that this effect occurs in addition to standard
  (i.e. IMF-invariant) variations of the relative number of young and
  old populations as a result of varying SFRs, making the empirical
  detection of IGIMF variations challenging.  We document in
Fig.~\ref{fig:maxmasfr} all quantities needed to construct, without
additional computation, IGIMFs for galaxies of all morphological types
and for a standard, a maximal and a minimal scenario. This ought to be
useful for future chemo-dynamical research within a hierarchical
cosmological structure formation scenario. 

In summary, given our assumptions the most important conclusions are:
\begin{itemize}
\item Chemical enrichment histories and SN rates calculated with an
  invariant Salpeter IMF may not be correct for any galaxy;
\item The number of supernovae is lower, and possibly significantly lower over
  cosmological times than for an invariant canonical IMF;
\item Irrespective of how old a galaxy is it will always appear less
  chemically evolved than a more massive equally-old galaxy as a
  result of the steeper IGIMF;
\item The scatter in chemical properties must increase with decreasing
  galaxy mass;
\item A steeper (input) IMF above $1\,M_\odot$ would aggravate the systematic
  differences in galaxy properties when compared to a Salpeter
  IMF as well as increasing the variations with galaxy mass;
\end{itemize}
where galaxy mass refers only to the assembled mass in stars.

 Of future interest would be a direct observational test of this
  model. This may be possible by using deep luminosity functions of
  nearby galaxies. But these functions will reflect the PDMF so that
  stellar evolution and the SFH need to be taken into account in order
  to extract the IGIMF.  The difficulty associated with such work is
  seen by several studies of starburst galaxies having found (IG)IMFs
truncated at the lower mass end near a few $M_{\odot}$
\citep[e.g.][]{WJR88}, but more recent studies \citep[][and references
  therein]{El04b,GBH04} showed that this seems not to be the case. In
the same paper Elmegreen points out that a Salpeter IMF with a
flattening below about 0.5 -- 1 $M_{\odot}$ gives a good approximation
for starburst galaxies. This does not necessarily contradict our
result of a variable galaxy-wide IGIMF, as the observable part of the
IGIMF in a starburst galaxy is dominated by a few young and massive
clusters for which we adopt a Salpeter IMF in the high mass range. In
our model variations come from the fainter less-massive clusters.
 
It should be noted here that for the more conservative approach
of the minimal scenario our results still predict considerable
differences in comparison with a constant canonical IMF. For example,
\citet{GP04} make assumptions rather similar to our minimal scenario
and predict a reduction of the general metal yield by a factor of about 1.8.


\section*{Acknowledgments}
We thank John Scalo for very useful suggestions. This work has been
funded by DFG grant KR1635/3.

\label{lastpage}

\end{document}